%% file: interleavedRegulator.tex
\documentclass[journal]{IEEEtran}
\usepackage{amsfonts}
\IEEEoverridecommandlockouts 
\usepackage{cite}
\ifCLASSINFOpdf
   \usepackage[pdftex]{graphicx}
   \graphicspath{{./Images/}}
   \DeclareGraphicsExtensions{.pdf,.jpeg,.png}
\else
\fi
\usepackage[cmex10]{amsmath}
\usepackage[font=footnotesize]{subfig}

\usepackage[hidelinks]{hyperref}
\usepackage{multirow}
\usepackage{booktabs,colortbl}

\input{./lebStuff}
\newtheorem{definition}{Definition}
\newtheorem{theorem}{Theorem}
\newtheorem{lemma}{Lemma}

\newcommand{\ntxt}[1]{#1} 

\begin{document}

\title{
A Theory of Traffic Regulators for Deterministic Networks with Application to Interleaved Regulators
}

\author{%

\IEEEauthorblockN{Jean-Yves~Le~Boudec}

%
}


\maketitle

\begin{abstract}

 \ntxt{We introduce Pi-regulation, a new definition of traffic regulation
which extends both the arrival curves of network calculus and Chang's max-plus g-regulation, and also includes new types of regulation such as packet rate limitations. We provide a new exact equivalence between min-plus and max-plus formulations of traffic regulation. We show the existence and a max-plus representation of per-flow minimal regulators, which extends the concepts of packetized greedy shapers and minimal g-regulators. We show that any minimal regulator, placed after any arbitrary system that is FIFO for the flow of interest, does not increase the worst-case delay of the flow. We extend the theory to interleaved regulation and introduce the concept of minimal interleaved regulator. It generalizes the Urgency Based Shaper that was recently proposed by Specht and Samii as a simpler alternative to per-flow regulators in deterministic networks with aggregate scheduling. With this regulator, packets of multiple flows are processed in one FIFO queue and only the packet at the head of the queue is examined against the regulation constraints of its flow. We show that any minimal interleaved regulator, placed after any arbitrary FIFO system does not increase the worst-case delay of the combination.
}

\end{abstract}


\input{intro}
\input{not}
\input{reg}
\input{pfr}

\input{ir}
\input{conc}

\bibliographystyle{unsrt}
\bibliography{leb}
\appendix
\input{app}
\end{document}

%% file: lebStuff.tex
\usepackage{lebListes}  
\noitemsep  

\usepackage{ifthen}

\usepackage{algorithm,algpseudocode}

\usepackage{color,pstcol,pstricks}
\newgray{backgroundgray}{0.90}
\definecolor{backgroundgray}{gray}{0.90}

\newenvironment{petit}
{\par\vspace{.5\baselineskip}\noindent\footnotesize}
{\nobreak\par\vspace{.5\baselineskip}}

\def\tempfolder{./temp/}

\providecommand{\imprimeTexteCache}{non}
\providecommand{\imprimeTexteSecret}{non}

\newif\ifsol
\newif\ifnfs

\ifthenelse{\equal{\imprimeTexteCache}{oui}}{\soltrue}{\solfalse }
\ifthenelse{\equal{\imprimeTexteSecret}{oui}}{\nfstrue}{\nfsfalse}

\ifsol
   \providecommand{\sol}[1]{\bs #1 \es}
\else
   \providecommand{\sol}[1]{}
\fi

\ifnfs
   \providecommand{\nfs}[1]{~\\{\footnotesize \sffamily \emph{\textbf{Note: } #1}~\\}}
\else
   \providecommand{\nfs}[1]{}
\fi

\newenvironment{mySolution}
{\red \it
\par\vspace{.5\baselineskip}\noindent\textbf{Solution.~}}
{\vspace{.5\baselineskip}}
\def\bs{\begin{mySolution}}
\def\es{\end{mySolution}}

\def\problemfile#1{%
\begin{filecontents}{\tempfolder#1}}

\ifnfs

\else

\fi

\ifnfs

\else

\fi

\newcommand{\mycaption}[1]{\caption{\small \sffamily #1}}

\ifnfs

\else

\fi

\ifnfs
\newcommand{\tbd}[1]{\textsc{To be done:\\#1}}
\else
\newcommand{\tbd}[1]{}
\fi

\newtheorem{maquestion}{\sc Question}[section]

\ifsol

\else

\fi

\newcommand{\imp}[1]{{\blue \bf #1}}

\newcommand{\bp}{ 
  \small \ttfamily
 \begin{tabbing}
 aaa\=aaa\=aaa\=aaa\=aaaaaaaa \= aaaaaaaaaa\= \kill
 }
\newcommand{\ep}{\end{tabbing}\normalfont\normalsize }

\newcommand{\fref}[1]{Figure~\ref{#1}}
\newcommand{\sref}[1]{Section~\ref{#1}}
\newcommand{\cref}[1]{Chapter~\ref{#1}}

\newcommand{\eref}[1]{Eq.(\ref{#1})}

\newcommand{\lref}[1]{Lemma~\ref{#1}}
\newcommand{\thref}[1]{Theorem~\ref{#1}}

\newcommand{\cqfd}{\hfill $\diamond$}

\newcommand{\calA}{\mathcal A}

\newcommand{\calF}{\mathcal F}
\newcommand{\calG}{\mathcal G}

\newcommand{\lp}{\left(}
\newcommand{\lb}{\left[}
\newcommand{\lc}{\left\{}

\newcommand{\rp}{\right)}

\newcommand{\rc}{\right\}}

\newcommand{\card}{\mbox{card}}

\newcommand{\nin}{\not\in} 

\newcommand{\ind}[1]{\mathbf{1}_{\{#1\}}}
\newcommand{\indinf}[1]{\mathbb{I}_{\{#1\}}}

\def\be{\begin{equation}}
\def\ee{\end{equation}}
\def\ben{\[}
\def\een{\]}
\def\bearn{\begin{eqnarray*}}
\def\eearn{\end{eqnarray*}}
\def\bear{\begin{eqnarray}}
\def\eear{\end{eqnarray}}
\def\barr{\begin{array}}
\def\earr{\end{array}}
\def\bel{\be \barr{l}}
\def\eel{\earr\ee}
\def\beln{\ben \barr{l}}
\def\eeln{\earr\een}

\def\bfig{\begin{figure}}
\def\efig{\end{figure}}
\def\bmat{\left(\begin{array}}
\def\emat{\end{array}\right)}

\newcommand{\limit}[2]{\lim_{#1 \rightarrow #2}}
\newcommand{\eqdef}{\stackrel{\mathrm{def}}{=}} 

\newcommand{\mand} {\mathrm{\;  and \; }}

\newcommand{\mst} {\mathrm{\;  such \;  that \; }}
\newcommand{\mfa} {\mathrm{\;  for \;  all \; }}
\newcommand{\mfor} {\mathrm{\;  for \;  }}

\def\Reals{\mathbb{R}}

\def\Nats{\mathbb{N}}

\newsavebox{\coloredbox}

\newsavebox{\traitbox}



%% file: intro.tex
\section{Introduction}

We are interested in First-In-First-Out (FIFO) or FIFO-per-class systems, as they are found \ntxt{e.g. in industrial networks or in} Ethernet or routed networks with delay guarantees, see for example the current time sensitive networking (TSN) group of IEEE 802.1 \cite{TSN} or the detnet group of IETF \cite{detnet}. In FIFO networks, the burstiness of a flow increases at every hop where it shares a queuing point with other flows of its class, in direct relation to the burstiness of these flows \cite{charny2000delay}.
The increased burstiness of this flow may, in turn, increase the burstiness of other flows in the same class. This creates a positive feedback loop, which causes large worst-case delays in FIFO per-class networks \cite{psrgton}. Another consequence is that computing worst-case delays in FIFO networks is a very difficult exercise~\cite{boyer2008tightening}. \ntxt{Even in feed-forward topologies, finding proven bounds that are close to the worst case is difficult: the determination of the worst case is NP-hard~\cite{5461912} and can be addressed by linear programming \cite{bouillard2015exact}. Alternative approaches combine the network calculus building blocks of min-plus network calculus in various ways, leading to better bounds by using the Pay Multiplexing Only Once property \cite{schmitt2008improving} or the Aggregate Arrival Bounding technique \cite{bondorf2016calculating}. However, none of these combinations seem to dominate the others and every case requires some tailored analysis \cite{bondorf2018catching}.  These techniques are available in several software tools \cite{bondorf2014discodnc,mifdaoui2010wopanets}.

A radical alternative is to avoid cascades of increased burstiness by re-shaping every flow at every hop \cite{ayed2014hierarchical}. This was used in tools that perform industry-floor automatisation by using network calculus \cite{kerschbaum2016need}. However, this typically requires per-flow queuing at every hop, which defeats the purpose of FIFO networks.
}

Specht and Samii introduced in \cite{specht2016urgency} a simpler alternative, under the name ``Urgency Based Scheduler" (UBS). With a UBS, the packet at the head of the queue is examined against the regulation constraints of its flow; it is released at the earliest time \ntxt{at which} this is possible without violating the constraints. Packets that are not at the head of the queue are not examined until they reach the head of the queue. The regulation constraints are either the  ``Length Rate Quotient" (LRQ) rule, or a leaky bucket constraint (we explain the details in \sref{sec-reg}).
The motivation is to avoid per-flow queuing, which is perceived as expensive, while keeping per-flow state, which is inexpensive. \ntxt{Similar ideas are used in the context of best-effort traffic in \cite{saeed2017carousel}.

In \cite{specht2016urgency}, Specht and Samii compute a bound on the packet delay for a network of priority queues with constant rate servers and UBS. They use a trajectory analysis. More precisely, end-to-end latency bounds are obtained by examining, for each flow, a sequence of past dequeuing events, and assuming that there are periods of times in the past where queues were empty. The analysis is ad-hoc, and does not separate between the service process and the effect of the UBS. It is very complex and is extremely difficult to validate. Nonetheless, it is quite remarkable, and somewhat unexpected, that Specht and Samii find explicit latency bounds, in spite of the interleaving of flows in the UBS. Also they find, by inspection of formulas, that UBS does not increase the delay bound that they obtain for the priority scheduler. The goal of this paper is to provide a theory that (1) can be used to explain and formally prove these results, (2) isolates the effect of the UBS and of the specific assumptions on the rest of the network, (3) applies to general FIFO systems instead of priority queues, and (4) separates what is specific to leaky bucket and LRQ and what is a consequence of minimal regulation in general. }

\ntxt{To this end}, we introduce the concept of ``Minimal Interleaved Regulator", which extends the concept of UBS to a very large class of regulation rules. Like the UBS, a minimal interleaved regulator examines only the packet at the head of the queue; it possibly delays this packet but also, due to FIFO, all following packets, which typically belong to other flows.  However, we show that, when a minimal interleaved regulator is placed after an arbitrary FIFO system, the worst-case delay of the combination is the same as without the interleaved regulator. This “shaping-for-free” property is well-known with per-flow shapers and per-flow service curve elements \cite{gue96}; \ntxt{as we show later, it also holds with general per-flow minimal regulators and arbitrary FIFO systems settings, and}  surprisingly, it continues to hold with minimal \emph{interleaved} regulators.

%

UBS assumes that flows are regulated either using LRQ or a leaky bucket constraint. Now the former is an instance of Chang's g-regulator \cite{Chang98}, which uses max-plus algebra (see \sref{sec-reg-g}), whereas the latter is an instance of an arrival curve constraint, which uses min-plus algebra (see \sref{sec-reg-ac}). It is known that a g-regulation constraint is not equivalent to an arrival curve constraint \cite{Changbook}.
In contrast, in \cite{liebeherr2017duality} Liebeherr shows that there is an isomorphism between the min-plus and max-plus representations of regulation constraints, and of the associated regulators (called ``shapers"). Therefore the non-equivalence between Chang's g-regulat\ntxt{ion} and arrival curve constraints is not explained by the choice a max-plus or min-plus algebra. Indeed, the difference is whether traffic constraints are expressed at an arbitrary packet arrival time (as with Chang's g-regulation constraint) or at an arbitrary point in time or space (as with arrival curves or the equivalent max-plus envelope of Liebeherr). \ntxt{In \thref{theo-mpmpgs} we give a new result which formulates arrival curve constraints by means of packet arrival times, and thus} clarifies the relation between these viewpoints.


This motivates the definition of a new formalism for regulators that encompasses both g-regulation constraints and arrival curves.
More precisely, we introduce
in \sref{sec-reg} the concept of ``Pi-regularity"  and in \sref{sec-pfr} of ``minimal $\Pi-$regulator",  adapted to the context of FIFO systems. We show that these concepts contain as special cases the classical greedy shaper of network calculus~\cite{lt01} 
as well as Chang's ``minimal regulator" \cite{Chang98}. 
We show that it also contains other regulators, which cannot be expressed in these classical frameworks, such as the TSN packet rate regulator. \ntxt{We show that any per-flow minimal regulator does not increase the worst-case latency of any system that is FIFO per flow. We also provide a simple, intuitive proof for this result.}


In \sref{sec-ir} we introduce and analyze interleaved regulators\ntxt{, show their existence and derive in an intuitive and simple way} our main result on the maximum latency induced by minimal interleaved regulators. 
\sref{sec-not} describes the notation and provides some background results. Proofs of lemmas and theorems are in appendix\ntxt{, except for the proofs of the theorems on worst-case latency, which have an independent interest, and are given in the main text}. 

%% file: not.tex
\section{Notation and General Pre-Requisites}
\label{sec-not}
\subsection{Packet Sequences and FIFO Systems}
\label{sec-not-1}
We use a notation similar to Chang's marked point process notation in \cite{Chang98}. We consider \ntxt{an arbitrary FIFO system that has} packet sequences as input and output. In some cases the input and output packet sequences  belong to one single flow, but in general we are interested in \ntxt{packet} sequences where packets may belong to different flows.

\begin{itemize}
  \item $\Nats=\lc 0, 1,2,...\rc$ and $\Nats^+=\lc1,2, 3,...\rc$.
  \item $\Reals^+=\lb 0, \infty\rp$ is the set of non-negative real numbers.
  \item $\calF$ is the set of sequences such that $A\in\calF$ whenever $A=(A_1, A_2, ...)$ with $A_n\in [-\infty, +\infty)$ for $n\in\Nats^+$. 
  \item
  $\calF_{inc}$ is the subset of $\calF$ of wide-sense increasing sequences, i.e. $A\in\calF_{inc}$ if and only if $A\in\calF$ and $A_n \leq A_{n+1}$ for all $n\in\Nats^+$.
  \item
  $\calG$ is the set of positive integer-valued sequences. i.e. $L\in\calG$ whenever $L=(L_1, L_2, ...)$ with $L_n\in \Nats^+$ for $n\in\Nats^+$.
  \item $x\vee y$ denotes the maximum of $x$ and $y$, for $x,y \in [-\infty, +\infty]$.
  \item Whenever $h\in \Reals$ and $A\in\calF$, $A+h$ and $h+A$ denote the sequence $A'$ such that $A'_n=A_n+h$ for all $n\in\Nats^+$.
  \item For $A, A'$ in $\calF$ the notation $A\leq A'$ means that $A_n\leq A'_n$ for all  $n\in\Nats^+$.
  \item The supremum of an empty set 
  is 
   $-\infty$.
   \item The infimum of an empty set 
  is 
   $\infty$.
  \item The summation of an empty set is $0$.
  \item $\ind{C}$ is equal to $1$ when the condition $C$ is true and is equal to $0$ otherwise.
  \ntxt{\item $\indinf{C}$ is equal to $1$ when the condition $C$ is true and is equal to $+\infty$ otherwise.}
  \item For $x\in\Reals$, $\lceil x  \rceil$ is the ceiling of $x$, namely the smallest integer $\geq x$;  $\lfloor x  \rfloor$ is the floor of $x$, namely the largest integer $\leq x$.
\end{itemize}

A general packet sequence is a triple such as $(A, L, F)$ where:
\begin{enumerate}
  \item The first element $A$ is the sequence of packet dates, i.e. the time instants $A=(A_1, A_2, ...)$ at which the packets are observed. We assume that packet numbering follows chronological order; simultaneous packet observation times in the same sequence are possible. Thus we require that $A\in\calF_{inc}$. Depending on the context, we may denote the sequence of packet dates with $A$ (for arrivals) or $D$ (for departures) etc.
  \item The second element is a sequence of packet lengths $L=(L_1, L_2, ...)$ with $L\in\calG$. Packet lengths are counted in some arbitrary data unit, typically in bytes or words of a fixed number of bytes.
  \item The third element is a sequence of flow numbers $F=(F_1, F_2, ...)$ with $F_n\in\Nats^+$, and $n\in\Nats^+$. In other words, $F_n=f$ means that packet $n$ belongs to flow $f$. To avoid cumbersome notation, we assume without loss of generality that the set of flow numbers is finite and that the subsequence of packets of flow $f$ is infinite, for every~$f$.
\end{enumerate}
With this notation,  we can express a FIFO system as a system that maps a given input packet sequence $(A, L,F)$ to an output packet sequence $(D, L,F)$ such that $A\leq D$. \ntxt{Observe that our definition of packet sequences allows simultaneous packet arrivals; inside a packet sequence, different packets with identical dates must be numbered differently, i.e., we assume that there is a tie-breaking rule for ordering packets. Our definition of FIFO system requires that such a numbering is preserved.}

When a packet sequence is for a single flow, we \ntxt{simply describe} it as a couple such as $(A, L)$. \ntxt{A system $S$ is FIFO for a flow with input sequence $(A,L)$ if the output packet sequence for this flow is $(D,L)$ for some $D \in\calF_{inc}$ such that $A\leq D$.}

\subsection{Notation for Flows Inside a Packet Sequence}
In \sref{sec-ir} we need some specific notation for flows inside a packet sequence.
%
%
Given a sequence $F$ of flow numbers, we define $I()$ as the function 
that returns the index of packet $n$ in its flow. In other words:
\be I(n)= \card \lc m\in \Nats^+ :   m \leq n  \mand F_m=F_n \rc
\ee

We also define the function $\mbox{ind}()$ such that $\mbox{ind}(f,i)$ is the index in the packet sequence of the $i^{th}$ packet of flow $f$; in other words:
\be
 \mbox{ind}(f,i) = n\Leftrightarrow \lp F_n=f \mand I(n)=i \rp
\ee


Note that the functions $I()$ and $\mbox{ind}()$ 
depend on the packet sequence $F$ but we leave out the dependency on $F$ for the sake of simplicity in notation.

When a flow $f$ is present in a packet sequence $(A, L, F)$ we define $A^f$ [resp. $L^f$] as the subsequence extracted from $A$ [resp. $L$] by keeping only the packet dates [resp. lengths] corresponding to a packet of flow $f$, namely
\be A^f_i=A_{\mbox{ind}(f,i)},\; L^f_i=L_{\mbox{ind}(f,i)} \mfa i\in\Nats\ee

For example, assume the sequence of flow numbers is $F=(3,4, 1,2, 1,3 ...)$, i.e. the first packet belongs to flow~$3$ ($F_1=3$), the second to flow~$4$ ($F_2=4)$, etc. Packets 3 and 5 belong to flow 1, packet $5$ is the second packet of flow $1$ so $\mbox{ind}(1,2)=5$, $I(5)=2$, $A^1=(A_3, A_5, ...)$ and $A^3=(A_1, A_6, ...)$.

\subsection{Pseudo-Inverses}
Let $f()$ be a wide-sense increasing function $\Reals^+\to \Reals^+$. Let $f^{\downarrow}():\Reals^+\to [0,\infty]$ be its lower pseudo-inverse, defined by \cite{Chang98,liebeherr2017duality}
\bear
f^{\downarrow}(x)&=&\inf\lc s \geq 0\mst f(s)\geq x\rc
\label{eq-inv-l1}
\\
&=&\sup\lc s \geq 0\mst f(s)< x\rc
\label{eq-inv-l2}
\eear
The lower pseudo-inverse is the same as the pseudo-inverse in \cite{lt01}. Similarly, let $f^{\uparrow}():\Reals^+\to [0,\infty]$ be its upper pseudo-inverse, defined by \cite{Chang98,liebeherr2017duality}
\bear
f^{\uparrow}(x)&=&\sup\lc s \geq 0\mst f(s)\leq x\rc\label{eq-inv-u1}\\
&=&
\inf\lc s \geq 0\mst f(s)> x\rc\label{eq-inv-u2}
\eear
Note that $f^{\downarrow}(0)=0$. Furthermore \cite[Theorem 3.1.2]{lt01}:
\be
f(t)\geq x \Rightarrow  t\geq  f^{\downarrow}(x)\label{eq-invineq0}
\ee
and \be
t> f^{\downarrow}(x)\Rightarrow f(t)\geq x \label{eq-invineq}
\ee but the converse may not hold. However, we have:
\begin{lemma}
If $f$ is right-continuous then for all $t,x \in \Reals^+$:
  \be
 t\geq f^{\downarrow}(x)   \Leftrightarrow f(t)\geq x
  \ee
\label{lem-leminv2}
\end{lemma}

It is known \ntxt{\cite[Theorem (4.2.1)]{dieudonne2013foundations}} that $f()$ is not necessarily continuous but (1) its set of discontinuities is countable and (2) $f()$ has a limit to the right and to the left at every point.
We denote with $f^+()$ the right-limit of $f()$, defined for $t \in \Reals^+$ by $f^+(t)=\limit{s}{t, s>t} f(s)=\inf_{s>t}f(s)$. Note that $f^+()$ is right-continuous and whenever $0\leq s<t$:
\be
 f(s)\leq f^+(s)\leq f(t)
 \label{eq-fplus2}
\ee
Similarly, we denote with $f^-()$ the left-limit of $f()$, defined for $t \in \Reals^+$ by $f\ntxt{^-}(t)=\limit{s}{t, s<t} f(s)=\sup_{s<t}f(s)$.

Last, we will use the following results, which are true for any wide-sense increasing function $f(): \Reals^+\to \Reals^+$
\begin{lemma}$f^{\downarrow}=\lp f^+ \rp^{\downarrow}$ and $f^{\uparrow}=\lp f^- \rp^{\uparrow}$.
\label{lem-inv}
\end{lemma}

\begin{lemma}${(f^{\uparrow})}^-=f^{\downarrow}$ and ${(f^{\downarrow})}^+=f^{\uparrow}$ .
\label{lem-inv-n}
\end{lemma}

 Proofs of the lemmas are in appendix.

%% file: reg.tex
\section{Pi-Regularity}
\label{sec-reg}
In this section and the next section we are interested in regulation of a single flow. We start by introducing a new concept, ``Pi-regularity", which extends both arrival curve constraints and Chang's g-regularity. This concept will prove to be essential in analyzing interleaved regulators in \sref{sec-ir}.

\subsection{Definition of Pi-Regularity}
Our new definition of regularity uses an operator, say $\Pi$, which must satisfy the following conditions.

\begin{description}
  \item[C1] $\Pi$ is a mapping $\calF_{inc}\times\calG\to\calF$, i.e. $\Pi$ takes as argument a single-flow packet sequence $(A,L)$ and transforms it into a sequence of time instants. The output sequence is in $\calF$, i.e. is not necessarily monotonic.
  \item[C2] \textbf{$\Pi$ is causal}\ntxt{, in the following sense:} if $\Pi(A,L)=A'$ then the value of $A'_n$ may depend on  $A_1,...A_{n-1}$ and $L_1, ...,L_n$ but not on $A_m$ for $m\geq n$ nor on $L_m$ for $m\geq \ntxt{n}+1$.
  \item[C3] \textbf{$\Pi$ is homogeneous with respect to $A$:} $\Pi(A+h, L)=\Pi(A, L)+h$ for any constant $h\in \Reals$ and any sequences $A\in \calF_{inc}, L\in \calG$.

  \item[C4] \textbf{$\Pi$ is isotone  with respect to $A$:} whenever $A,A'\in \calF_{inc}$ are such that $A\leq A'$ then also  $\Pi(A, L)\leq \Pi(A', L)$ for any sequence $L\in\calG$.
\end{description}
%
%
\ntxt{Observe that the causality condition C2 is a little unusual, as it does not allow $A'_n$ to depend on $A_n$. This is required for the theory to work; in particular, if we would allow $A'_n$ to depend on $A_n$, the max-plus representations of the minimal regulators in \eref{eq-defminreg} and \eref{eq-defminir} would be circular and would not be useful.}

Note that if an operator $\Pi$ is causal and homogeneous with respect to $A$ then necessarily $\Pi(A, L)_1=-\infty$ for any input $(A, L)$. This can be derived by observing first that $\Pi(A, L)_1$ is independent of $A$ by causality and therefore has the form $c(L)\in [-\infty, \infty)$. Second, by homogeneity $\Pi(A+h, L)=\Pi(A, L)+h$ for any real number $h$ so $c(L)=c(L)+h$ for any $h$, which is possible only if $c(L)=-\infty$.

In the next three subsections we give several examples of such operators. They all have the form
\be
\Pi(A, L)_n=\max_{1\leq m\leq n-1}\lc A_m + H_{m,n}(L) \rc
\label{eq-mpo}
\ee for some appropriate choice of the array $H_{m,n}(L)$ (with $H_{m,n}(L)\in [-\infty, \infty)$ for $m \ntxt{<} n\in\Nats^+$, $L\in\calG$ \ntxt{ and such that $H_{m,n}$ does not depend on $L_j$ for $j\geq n+1$}).  An operator defined by an equation of the form \eref{eq-mpo} is a max-plus-linear operator and clearly satisfies C1-C4. Note that here the identity $\Pi(A)_1 =-\infty$ follows from the fact that the max \ntxt{in \eref{eq-mpo} is $-\infty$ when the set of indices is empty. }

\begin{definition}[Pi-Regularity.]
Given some operator $\Pi$ that satisfies C1-C4, we say that a single-flow packet sequence $(A, L)$ is $\Pi-$regular if
$ A\geq \Pi(A, L)$.
\label{def-pi}
\end{definition}

We next give some examples and show how this definition extends existing frameworks.
\subsection{Chang's g-Regularity}
\label{sec-reg-g}
Given some \ntxt{wide-sense increasing} function (or sequence) $g(): \Nats\to\ntxt{\Reals^+}$ \ntxt{such that $g(0)=0$}, Chang \cite{Chang98} defines a single-flow packet sequence $(A,L)$ as $g$-regular if and only if for all $m,n\in \Nats^+$ such that $m<n$ we have
\be
A_n-A_m\geq g\lp L_m+...+L_{n-1}\rp
\label{eq-changreg}
\ee
It is immediate to see that g-regularity is a special case of Pi-regularity with the operator $\Pi$ given by
\be
\Pi(A,L)_n=\max_{1\leq m\leq n-1}\lc A_m + g\lp \sum_{j=m}^{n-1} L_j\rp \rc
\label{eq-chang-g}
\ee
In the case where $g(x)=x/r$ for some $r$, g-regularity is called the ``Length Rate Quotient" (LRQ) constraint with rate $r$ in \cite{specht2016urgency}. Any flow that is observed on a physical communication link of rate $r$ satisfies the LRQ($r$) constraint. Because in this case $g$ is linear, it can easily be seen that, for the LRQ($r$) regulation constraint, \eref{eq-changreg} is equivalent to the simpler condition
\be
A_n-A_{n-1}\geq \frac{L_{n-1}}{r}
\label{eq-lrq}
\ee for all $n\in\Nats^+$, $n\geq 2$. Therefore, LRQ($r$) regularity is an instance of Pi-regularity with the operator $\Pi^{LRQ(r)}$ given by
\bel
\Pi^{LRQ(r)}(A,L)_n= A_{n-1} + \frac{L_{n-1}}{r}\mfor n\geq 2
\label{eq-pi-lrq}
\\
\Pi^{LRQ(r)}(A,L)_1=-\infty
\eel

\subsection{Arrival Curve Constraint}
\label{sec-reg-ac}
This is a classical network calculus constraint, originally expressed with min-plus algebra. It uses some wide-sense increasing function $\sigma(): \Reals^+\to \Reals^+$, called ``arrival curve" or ``min-plus traffic envelope"; it can always be assumed without loss of generality that $\sigma$ is sub-additive and $\sigma(0)=0$ 
(but we don't need such an assumption in the following theorems). The celebrated leaky bucket constraint LB($r,b$) \cite{cru91a} corresponds to $\sigma(t)=rt+b$ for $t>0$ and $\sigma(0)=0$, where $r$ is the leaky bucket rate and $b$ the burstiness. The arrival curve constraint is expressed in terms of the cumulative arrival function $R(t)$, defined for $t\geq 0$, and which can be derived from the single-flow packet sequence $(A,L)$ by
\be
R(t)=\sum_{n\in\Nats^+}L_n \ind{A_n<t}
\label{eq-defr}
\ee
where the indicator function $\ind{A_n<t}$ has the value $1$ when the condition $\lc A_n<t\rc$ is true and $0$ otherwise. 
In this context, we assume that time is nonnegative and in particular $A_n\geq 0$. The arrival curve constraint requires that
\be
R(t)-R(s)\leq \sigma(t-s)\mfa 0\leq s\leq t \label{eq-ac}
\ee

Liebeherr shows in\cite{liebeherr2017duality} that the arrival curve constraint can be expressed in max-plus algebra. To this end, he introduces the arrival time function $ T()=R^{\uparrow}()$, which is equal to the upper pseudo-inverse of cumulative arrival function $R()$, and is also given by
\be
T(x)= \inf_{n\in\Nats^+}\lp A_n \ntxt{\indinf{L_1+...+L_n>x}}\rp \mfor x\geq 0
\label{eq-deft}
\ee
\ntxt{where the indicator function $\indinf{L_1+...+A_n>x}$ has the value $1$ when the condition $\lc A_n<t\rc$ is true and $+\infty$ otherwise.}
Note that $R()$ can be recovered \ntxt{from} $T()$ since $R()=T^{\downarrow}()$. Liebeherr shows that the arrival curve constraint \eref{eq-ac} is equivalent to the condition
\be
 T(y)-T(x)\geq \lambda(y-x) \mfa 0\leq x\leq y \label{eq-mpte}
\ee
with $\lambda()=\sigma^{\uparrow}()$. In other words, the upper pseudo-inverse $\sigma^{\uparrow}()$ is a max-plus traffic envelope of $T()$ if and only if $\sigma()$ is an arrival curve, or min-plus traffic envelope, of $R()$. If we enforce that $\sigma()$ be left continuous, then $\sigma()$ is the lower pseudo-inverse of $\sigma^{\uparrow}()$ and thus there is exact equivalence between min-plus and max-plus traffic envelopes.

To make the link between arrival curve constraints and Pi-regularity, we need to go one step further and understand the relation between traffic constraints expressed at an arbitrary point in time (or space, as in \eref{eq-ac} and \eref{eq-mpte}) and constraints that are expressed at packet arrival times. This is provided by the following theorem.




\begin{theorem}\ntxt{[Formulation of Arrival Curve Constraint by Means of Packet Arrival Times]}
\label{theo-mpmpgs}
Consider a single-flow packet sequence $(A,L)$ and the associated cumulative arrival function $R()$ given by \eref{eq-defr}. Let $\sigma()$ be some wide-sense increasing function, $\sigma^+()$ its right-limit  and $\sigma^{\downarrow}()$ the lower pseudo-inverse of $\sigma()$ (and hence, by \lref{lem-inv} also of $\sigma^{+}()$).

 The following three conditions are equivalent:
\begin{enumerate}
  \item The arrival curve constraint in \eref{eq-ac} is satisfied;
  \item For any $m,n$ with $1\leq m\leq n$:
  \be
  \sum_{j=m}^{n} L_j\leq \sigma^+(A_n-A_m)\label{eq-mpmpg2}
  \ee
  \item For any $m,n$ with $1\leq m\leq n$: %
  \be
   A_n-A_m \geq \sigma^{\downarrow}\lp\sum_{j=m}^{n} L_j\rp\label{eq-mpmpg4}
\ee
\end{enumerate}
\end{theorem}

The proof is in appendix.
We give above a version of the theorem using the arrival curve $\sigma()$. An equivalent formulation can be given, assuming that we are given not the arrival curve $\sigma()$ but the max-plus traffic envelope $\lambda()$. Then define $\sigma=\lambda^{\downarrow}$ so that \eref{eq-ac} is equivalent to \eref{eq-mpte}. Using \lref{lem-inv-n} we have $\lambda^{\uparrow}=\sigma^+$ and $\lambda^{-}=\sigma^{\downarrow}$, so that \eref{eq-mpmpg2} is equivalent to
\be
\sum_{j=m}^{n} L_j\leq \lambda^{\uparrow}(A_n-A_m)
\label{eq-mpmpg2a}\ee  and \eref{eq-mpmpg4} is equivalent to
\be A_n-A_m \geq \lambda^{-}\lp\sum_{j=m}^{n} L_j\rp
\label{eq-mpmpg4a}
\ee
which shows the duality between min-plus and max-plus representations.

\ntxt{Theorem~1 shows that an arrival curve constraint, which is originally defined by using cumulative arrival functions, can also be expressed \emph{exactly} by using packet arrival times. Thus, its establishes an exact equivalence between a min-plus oriented representation, and a max-plus oriented representation.

An
 important outcome} is that the condition in
\eref{eq-mpmpg4} or \eref{eq-mpmpg4a} has the form in \eref{eq-mpo}; this shows that an arrival curve constraint is a special form of Pi-regularity. The operator $\Pi$ that corresponds
to an arrival curve $\sigma()$ is given by
\be
\Pi(A,L)_n = \max_{1\leq m \leq n-1} \lc A_m + \sigma^{\downarrow}\lp\sum_{j=m}^{n} L_j\rp\rc
\label{eq-pi-ac}
\ee

The theorem also \ntxt{sheds some light on the difference} between arrival curve constraints and Chang's g-regularity. Notice that the arrival curve constraint in \eref{eq-ac} implies the condition
\be
\sum_{j=m}^{n-1} L_j\leq \sigma(A_n-A_m),
\label{eq-mpmpg1}
\ee for any $m\leq n\in\Nats^+$. This can be derived from a direct application of \eref{eq-ac} to $s=A_m, t=A_n$.
However, it can easily be seen that the converse is not true, i.e. this last condition is not equivalent to the arrival curve constraint -- compare to \eref{eq-mpmpg2}. Observe now that
the definition of g-regularity involves only equations such as
\eref{eq-mpmpg1}. In particular, it does not involve the length $L_n$ of the current packet, which explains why it cannot be equivalent to arrival curve
constraints and why published relations
between the two involve a bound with a term in $L^{\max}$, the maximum packet size of the flow.

In the rest of this subsection we apply the above theorem to two classic arrival curve constraints.

\subsubsection{Leaky Bucket Constraints}
For the single leaky bucket constraint LB($r,b$), we can take $\sigma(t)=rt+b$ for some positive $r$ and $b$, so that $\sigma^+(t)=\sigma(t)$ and $\sigma^{\downarrow}(x)=0 \vee \frac{x-b}{r}$. The condition \eref{eq-mpmpg4} is equivalent to
\be
A_n\geq A_m\vee \frac{\lp\sum_{j=m}^{n} L_j\rp-b}{r}
\ee
and the Pi-regularity condition can be expressed by
\bear
 A_n&\geq& A_1\vee \frac{\lp\sum_{j=1}^{n} L_j\rp-b}{r}\vee A_2\vee \frac{\lp\sum_{j=2}^{n} L_j\rp-b}{r}\nonumber \\
 &&...\nonumber\\
 &&\vee A_{n-1}\vee \frac{\lp\sum_{j=n-1}^{n} L_j\rp-b}{r}
\eear
Note that $A_n\geq A_m$ is always true by construction for any $1\leq m \leq n-1$. Therefore, the constraint in the previous equation is equivalent to
\be
A_n\geq \max_{1\leq m\leq n-1}\lc A_m + \frac{\lp\sum_{j=m}^{n} L_j\rp-b}{r}\rc
\label{eq-lb7}
\ee
The operator $\Pi^{LB(r,b)}$ that corresponds to the leaky bucket constraint LB($r,b$) is therefore given by

\be
\Pi^{LB(r,b)}(A,L)_n = \max_{1\leq m \leq n-1} \lc A_m + \frac{\lp\sum_{j=m}^{n} L_j\rp-b}{r}\rc
\label{eq-pi-lb}
\ee

\subsubsection{Staircase Arrival Curve}

The staircase arrival curve SC($\tau, b$) is defined for $\tau>0, b>0$ by
\be
\sigma(t)=b \left\lceil \frac t\tau\right\rceil, \; t\geq0
\ee and is used to express the constraint that at most $b$ data units can be observed over any window of \ntxt{fixed} duration $\tau$~\cite{lt01}. Here $\sigma()$ is left-continuous and straightforward computations give:
\bear
\sigma^+(t)&=&b \left\lfloor \frac t\tau +1 \right\rfloor, \; t\geq 0
\\
\sigma^{\downarrow}(x)&=&\tau \left\lceil \frac x b - 1\right\rceil , \; x>0\label{eq-sc}
\\
\sigma^{\downarrow}(0)&=&0
\eear
The operator $\Pi^{SC(\tau, b)}$ that corresponds to the staircase arrival curve SC($\tau, b$) is therefore given by

\be
\Pi^{SC(\tau, b)}(A,L)_n = \max_{1\leq m \leq n-1} \lc A_m +\tau
\left \lceil
\frac{\lp\sum_{j=m}^{n} L_j\rp-b}{b}
\right \rceil \rc
\label{eq-pi-sc}
\ee

%
%

\subsection{Limits on Packet Rate}
\label{sec-lpr}

Packet rate limitations are used to put limits on the processing demand in networking boxes. For example, the IEEE TSN working group specifies a limit, say  $K\in \Nats^+$ on the number of packets sent by a flow over a specified duration, say $\tau\geq 0$. If all packets are of the same size, say $\ell$, this is the same as a staircase arrival curve constraint SC($\tau, b$) with $b=K\ell$. By \thref{theo-mpmpgs} and \eref{eq-sc}, in this specific cases such a packet rate constraint is equivalent to

 \be
 A_n-A_m\geq \tau\left\lceil
  \frac{n-m+1-K}{K}
 \right\rceil
 \label{eq-prsc}
 \ee

If packets are not all of the same size, this cannot be specified exactly as an arrival curve constraint nor as a g-regulation constraint. However, it is still expressed by the constraint in \eref{eq-prsc}. Therefore, this ``TSN packet rate regulation" is an instance of Pi-regularity, with the operator $\Pi^{TSN(\tau, K)}$ defined by

 \be
\Pi^{TSN(\tau, K)}(A,L)_n = \max_{1\leq m \leq n-1} \lc A_m +\tau
\left \lceil
\frac{n-m+1-K}{K}
\right \rceil \rc
\label{eq-pi-prsc}
\ee

Other forms of packet rate limitations can be defined. For example, a packet spacing constraint PS($\tau$) can be defined by
\be
A_n- A_{n-1}\geq \tau, \; \mfa n=2,3...
\ee
where $\tau\geq0$ is the spacing interval.
Note the analogy with an LRQ regulation constraint. This regulation constraint, is obviously an instance of Pi-regularity, with the operator $\Pi^{PS(\tau)}$ given by
\bel
\Pi^{PS(\tau)}(A,L)_n= A_{n-1} + \tau \mfor n\geq 2
\label{eq-pi-ps}
\\
\Pi^{PS(\tau)}(A,L)_1=-\infty
\eel

Last, similar to leaky bucket constraints, we can define a packet rate constraint by allowing a packet rate limit $\rho$ with some packet burstiness constraint $K$. In other words, the ``packet burstiness" constraint PB($\rho, K$) specifies that the number of packets observed over an interval of duration $t$ must be upper bounded by $\rho t +K$. \ntxt{A benefit of this formulation is its superposition property, i.e. if every flow $f$ in a set $S$ of flows is PB($\rho_f, K_f$) constrained, then the superposition is PB($\rho, K$) constrained, with $\rho=\sum_{f\in S} \rho_f $ and $K=\sum_{f\in S} K_f $. This follows immediately from the definition.   }

Similarly to \eref{eq-lb7}, the PB($\rho, K$) constraint can be expressed by the condition
\be
A_n\geq \max_{1\leq m\leq n-1}\lc A_m + \frac{n-m+1-K}{\rho}\rc
\label{eq-pb-mp}
\ee
\ntxt{In \cite{jiang2018basic}, Jiang says that a flow is $(\lambda,\nu)$ constrained if for all $m< n$
\be
A_n-A_m\geq \frac{1}{\lambda}\lp n-m-\nu\rp
\ee
It is straightforward to see that this is equivalent to \eref{eq-pb-mp} with $\lambda=\rho$ and $\nu=K-1$, in other words, Jiang's $(\lambda,\nu)$ constraint is the same as the PB($\lambda, \nu+1$) constraint.}

It follows \ntxt{from \eref{eq-pb-mp}} that the packet \ntxt{burstiness} constraint \ntxt{(hence also the $(\lambda,\nu)$ constraint) }is an instance of Pi-regularity, with the operator $\Pi^{PB(\rho,K)}$ given by

\be
\Pi^{PB(\rho,K)}(A,L)_n = \max_{1\leq m \leq n-1} \lc A_m + \frac{n-m+1-K}{\rho}\rc
\label{eq-pi-pb}
\ee

Obviously, all of the constraints defined in this section can neither be expressed with arrival curves nor with g-regularity.

\subsection{Combination of Regulation Constraints}
Pi-regulation constraints can easily be combined, by taking the maximum of the operators. Indeed, it immediately follows from  Definition~\ref{def-pi} that a flow $(A,L)$ is both $\Pi^1$ and $\Pi^2$-regular if and only if it is $\Pi$-regular, with $\Pi$ being the maximum of $\Pi^1$ and $\Pi^2$, defined by
\be
\Pi(A,L)_n=\Pi^1(A,L)_n\vee\Pi^2(A,L)_n
\ee
\ntxt{for all $n\in\Nats^+$.}
It is straightforward to verify that if $\Pi^1$ and $\Pi^2$ both satisfy C1-C4, then so does $\Pi$.

%% file: pfr.tex
\section{Per-Flow $\Pi-$Regulator}
\label{sec-pfr}
After defining Pi-regularity we can now define a per-flow ``$\Pi$-regulator" as a FIFO system that may delay some or all of the packets of a flow in order to make sure that the resulting output is $\Pi$-regular, for some operator $\Pi$. A  \emph{minimal} $\Pi$-regulator is one that outputs the packets of the flow as early as possible. Its existence is shown next.

\subsection{Minimal Per-Flow $\Pi-$Regulator}
\begin{theorem}
 Consider a single-flow packet sequence $(A,L)$ and let $\Pi$ be an operator that satisfies C1-C4. The ``minimal $\Pi$ regulator" is defined as the FIFO system that transforms the input packet sequence $(A,L)$ into the output packet sequence $(D,L)$ such that $D_1=A_1$ and
\be
D_n = \max\lc A_n, D_{n-1},\Pi(D,L)_{n}\rc
\label{eq-defminreg}
\ee
\begin{enumerate}
  \item The system defined in this way is a $\Pi$-regulator for this flow.
  \item (Minimality:) For any other $\Pi$-regulator that transforms $(A,L)$ into say $(D', L)$ we have
$D'_n\geq D_n$ for all $n\in\Nats^+$.
 \item The flow $(A,L)$ is $\Pi$-regular if and only if $D=A$.
\end{enumerate}
  \label{theo-minreg}
\end{theorem}
The proof is in appendix. Note that the definition in \eref{eq-defminreg} may appear to be circular, but it is not. This is because $\Pi$ is assumed to be causal as in condition C2, which implies that $\Pi(D,L)_n$ depends only on $D_1,...,D_{n-1}$. Therefore, the output sequence $D$ is well defined by the initial condition $D_1=A_1$ and \eref{eq-defminreg}.

To each of the regulators described in the previous section is thus associated a corresponding minimal per-flow regulator, whose input\ntxt{-}output equation is given by \eref{eq-defminreg}. Sometimes \ntxt{this} equation can be simplified \ntxt{by removing redundant terms in the maximum operation or in the expansion of $\Pi(D,L)_{n}$; for example, for the packet spacing constraint PS($\tau$), \eref{eq-defminreg} becomes} $
D_n = A_n\vee D_{n-1} \vee (D_{n-1}+\tau)$, which, because $\tau\geq 0$, can be written more simply as
\be
D_n = A_n\vee (D_{n-1}+\tau)
\ee

\ntxt{We next }show 
that Chang's minimal regulator and the classic packetized greedy shaper \cite{LeBoudec2002mAugust} are all special cases of minimal per-flow $\Pi$-regulators. 

\subsection{Chang's Minimal Regulator, LRQ($r$)-Regulator.}
Chang's minimal $g$-regulator is the minimal FIFO system that delivers an output that is $g$-regular \cite{Chang98}. We have seen in \sref{sec-reg-g} that g-regularity is equivalent to Pi-regularity, with $\Pi$ given by \eref{eq-chang-g}. Therefore, the minimal $g$-regulator is the minimal $ \Pi$-regulator with $\Pi$ given by \eref{eq-chang-g}.

In particular for the minimal LRQ($r$)-regulator, a simpler formulation of the operator $\Pi$ is given by \eref{eq-pi-lrq}, from where we derive the input-output equations
$
D_n = \max\lc A_n,D_{n-1},D_{n-1} + \frac{L_{n-1}}{r}\rc
$.
Observe that $\frac{L_{n-1}}{r}\geq 0$ so the term $D_{n-1}$ can be removed from the max. The input-output equations of the minimal LRQ($r$)-regulator are thus $D_1=A_1$ and for $n\geq 2$:
\bel
 D_n= \max\lc A_n,D_{n-1} + \frac{L_{n-1}}{r}\rc
 \label{eq-lrq-reg}
 \eel

 \subsection{Packetized Greedy Shaper.}
Given some arrival curve $\sigma()$, a packetized shaper is a system that delivers an output that is packetized and satisfies the arrival curve constraint $\sigma()$. Among all packetized shapers, the packetized \emph{greedy} shaper is the one that delivers its output as early as possible \cite{LeBoudec2002mAugust}. The input and output of a packetized shaper are cumulative arrival functions such as $R()$ in \eref{eq-defr}. Given a sequence of packet lengths $L$ and for inputs and outputs that are packetized, there is a one-to-one mapping between the $R()$ function and the sequence of packet arrival times. Furthermore, by \thref{theo-mpmpgs}, an arrival curve constraint is a form of Pi-regularity, with $\Pi$ given by \eref{eq-pi-ac}. It is then straightforward to see that a packetized greedy shaper for the arrival curve $\sigma()$ is the same as the minimal $\Pi$-regulator with $\Pi$ given by \eref{eq-pi-ac}.

In the special case of a single leaky bucket constraint LB($r,b$), the packetized greedy shaper is called the leaky bucket shaper. Here the $\Pi$ operator can take the simpler form in \eref{eq-pi-lb}. The input\ntxt{-}output equations of the leaky bucket shaper LB($r,b$) are thus $D_1=A_1$ and for $n\geq 2$:
\bear
 D_n&=& A_n \vee D_{n-1} \vee\lp A_m + \frac{\lp\sum_{j=m}^{n} L_j\rp-b}{r}\rp\nonumber\\
  &&\vee ...\vee \lp A_{n-1} + \frac{\lp\sum_{j=n-1}^{n} L_j\rp-b}{r}\rp
 \label{eq-lb-reg}
 \eear
The relation above is the max-plus equation of a leaky bucket shaper. It is not the best formula for a practical implementation (see \cite{LeBoudec2002mAugust} for a discussion of the different implementations of leaky bucket shapers and their combinations) but it can be used to derive formal properties of such shapers, as we do in \sref{sec-ir}. 

\ntxt{

\subsection{Minimal Per-Flow Regulator Does Not Increase Worst-Case Delay}
Regulators are used as a means to avoid burstiness cascades. However, they also add some delay, which needs to be accounted for, when computing end-to-end delay bounds. We show next a ``reshaping-for-free" property, which is reminiscent of a similar property of networks that use per-flow queuing \cite{gue96}. 

More precisely, assume that a flow $(A,L)$ is fed into a system that is FIFO for this flow (\fref{fig-f1pf}). This system can be for example a localized system such as a queuing point inside a switch, router or end-system, which handles multiple flows and respects the order of packets for this flow; or it can be an entire network path that is FIFO for this flow. Assume that, at the input, the flow is $\Pi$-regular. This property is, generally, lost by the output flow $(D,L)$. Assume that we place a \emph{minimal} regulator for this flow, just after the output, in order to re-create the regularity that was lost by traversing the FIFO node and let $(E,L)$ be the reshaped flow, i.e., the output sequence of the minimal regulator. The following theorem establishes that the minimal regulator does not add anything to the worst-case delay of the FIFO system.

\begin{figure}[h]
 \includegraphics[width=0.50\textwidth]{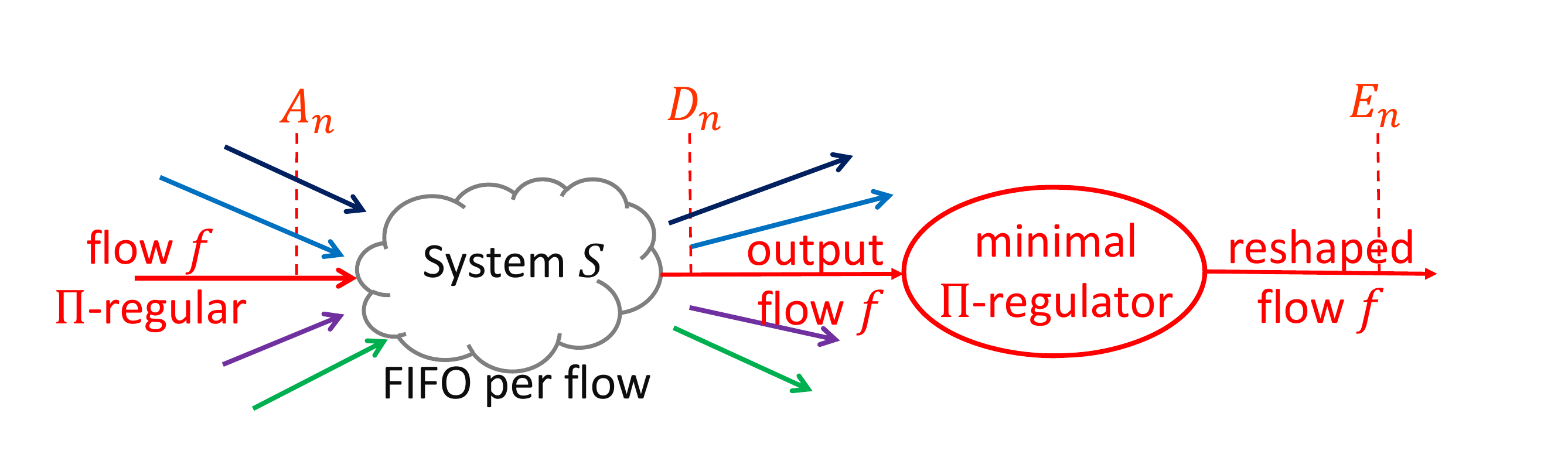}
\mycaption{Configuration for \thref{theo-pfr}. }
\label{fig-f1pf}
\end{figure}

\begin{theorem}
\label{theo-pfr}
  Assume one $\Pi$-regular flow is input to a system $S$ that is FIFO for this flow. Assume that the operator $\Pi$ satisfies conditions C1-C4. The output flow is fed to a minimal  regulator with same operator $\Pi$. The worst-case delay experienced by this flow through the combination is the same as the worst-case delay experienced through system $S$ alone.

  In other words, with the notation above:
\be
\sup_{n\in\Nats^+}\lp D_n-A_n\rp = \sup_{n\in\Nats^+}\lp E_n-A_n\rp
\label{eq-sff-pf}
\ee
\end{theorem}
The proof is illuminating and is given next. It shows that the theorem immediately derives from the minimality of the regulator.
\paragraph*{Proof} Let $d$ be the worst-case delay for this flow at system $S$. If $d=+\infty$ the conclusion is trivially true, thus we can assume that $d$ is finite.

Replace the minimal $\Pi$-regulator with a \emph{damper} with parameter $d$ \cite{verma1991guaranteeing}, as in \fref{fig-f1prf}. The damper, with parameter $d$, is a (theoretical) device that knows the time of arrival $A_n$ for every packet $n$ and delivers it at time $A_n+d$. The damper input is the flow $(D,L)$ therefore the damper is a system that is FIFO for this flow, as defined in \sref{sec-not-1}, if $D_n\leq A_n+d$, which holds because $d$ is an upper bound to the delay of this flow through $S$. The output of the damper is $(d+A, L)$; by property C3, it is also $\Pi$-regular. Therefore, the damper is a $\Pi$-regulator for this flow. By definition of the minimal $\Pi$-regulator, it follows that $E_n\leq A_n+d$.

\begin{figure}[h!]
 \includegraphics[width=0.50\textwidth]{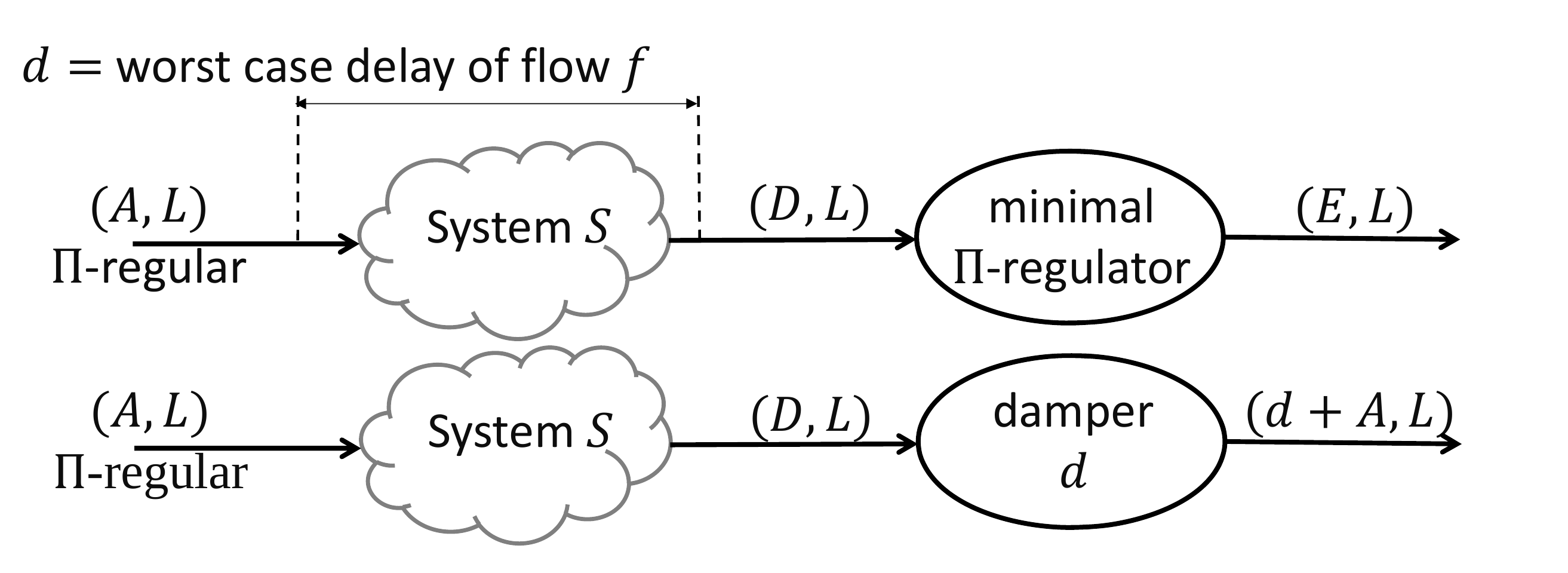}
\mycaption{Proof of \thref{theo-pfr}. }
\label{fig-f1prf}
\end{figure}
\cqfd

\thref{theo-pfr} extends of a well-known result concerning packetized greedy shapers, which states that re-shaping does not increase the delay bound provided by service curve elements. Here, our result is more general as it applies to any minimal regulator, not just to packetized greedy shapers.
}

%% file: ir.tex
\section{Interleaved Regulator}
\label{sec-ir}

In this section we consider a packet sequence $(A,L,F)$ of several multiplexed flows. Recall here that $L_n$ is the length of the $n^{th}$ packet, which belongs to flow $F_n$. Assume that for every flow $f$ we have a regulation constraint, with operator $\Pi^f$ for flow $f$. The regulation operators $\Pi^f$ may be of any kind, as long as they satisfy the conditions C1-C4. For example, the regulation operator may be of the LRQ type, leaky bucket or packet rate limit, or any combination of these operators. Furthermore, the regulation operators of different flows need not be of the same type. Recall that saying that flow $f$ is $\Pi^f$-regular means here that $A^f\geq \Pi^f(A^f, L^f)$ where $A^f, L^f$ are the sequences extracted from $A$ and $L$ by keeping only the indices corresponding to packets of flow $f$.

In this context we define an ``interleaved regulator" as a system that is FIFO and may delay some or all of the packets of the input sequence so that every flow $f$ inside the output sequence is $\Pi^f$-regular. Note that the FIFO condition imposes that a packet of a flow may not be delivered before a packet of some other flow that arrived before it. Formally, given a collection of regulation operators $\Pi^f$ \ntxt{that satisfy C1-C4}, with one operator per flow, an interleaved regulator is a FIFO system that transforms an input sequence $(A,L,F)$ into an output sequence $(D,L,F)$ such that $(D^f, L^f)$ is $\Pi^f-$regular for every flow $f$.

\subsection{Minimal Interleaved Regulator}
The following results establishes the existence of a minimal interleaved regulator, i.e. one that delays the packets as little as possible.
\begin{theorem}
Consider a packet sequence $(A,L,F)$ with, for every flow $f$, one regulation operator $\Pi^f$ that satisfies C1-C4. The ``minimal interleaved regulator" is defined as the FIFO system that transforms the input packet sequence $(A,L,F)$ into the output packet sequence $(D,L,F)$ defined by $D_1=A_1$ and
\be
D_n = \max\lc A_n, D_{n-1},\Pi^{F_n} \lp D^{F_n}, L^{F_n}\rp_{I(n)}\rc
\label{eq-defminir}
\ee
Recall that, in the above formula, $I(n)$ is the index of packet $n$ in its flow (namely in flow $f=F_n$).
\begin{enumerate}
  \item The system defined in this way is an interleaved regulator for this packet sequence.
 \item (Minimality:) For any other interleaved regulator that transforms $(A,L,F)$ into say $(D', L,F)$ we have
$D'_n\geq D_n$ for all $n\in\Nats^+$.
\item Every flow $f$ in $(A,L,F)$ is $\Pi^f$-regular if and only if $D=A$.
\end{enumerate}
  \label{theo-ir-min}
\end{theorem}

The proof is in appendix. \eref{eq-defminir} is the input-output characterization of the minimal interleaved regulator. It has an important consequence: it shows that the minimal interleaved regulator can be implemented as a ``head of the line" system, as in \cite{specht2016urgency}. More precisely, a possible implementation of the minimal interleaved regulator is as follows.
\begin{itemize}
  \item Packets of the multi-flow sequence are queued in FIFO order;
  \item The packet at the head of the queue is examined against the regulation constraints of its flow; it is released at the earliest time where this is possible without violating the constraints;
  \item Packets that are not at the head of the queue are not examined until they reach the head of the queue.
\end{itemize}

The Urgency Based Scheduler of Specht and Samii in \cite{specht2016urgency} is an instance of minimal interleaved regulator, which corresponds to the case where the regulation operator $\Pi^f$ is either of the form $\Pi^{LRQ(r_f)}$ as in \eref{eq-pi-lrq} or $\Pi^{LB(r_f,b_f)}$ as in \eref{eq-pi-lb}.

Since the minimal interleaved regulator uses a FIFO queue, there is no need for per-flow queuing. Note that there is per-flow state, but on one hand, this per-fow state can be very simple (a single number) for simple regulation rules such as leaky bucket, LRQ, packet spacing \ntxt{or packet burstiness}; on the other hand, per-flow state is typically present in switches and routers for packet forwarding and the per-flow state of the interleaved regulator can be placed there. In contrast, implementing one queue per flow as would be required by per-flow regulators has considerably larger complexity.

It follows from the structure of the minimal interleaved regulator that, when the packet at the head of the queue is not eligible for delivery, all packets behind it are delayed. Therefore a packet may be delayed either because it is too early with respect to the regulation imposed to its flow, or because a packet of some other flow at the head of the \ntxt{queue} is being delayed.

\subsection{Minimal Interleaved Regulator Does Not Increase Worst-Case Delay}
As discussed in the introduction, interleaved regulators are used in networks that handle multiple flows in the same queue, as a means to avoid burstiness cascades. However, they also add some delay, which needs to be accounted for\ntxt{, when computing} end-to-end delay bounds.
\ntxt{However, they enjoy the same property as their per-flow counterparts, which is also a consequence of their minimality.} 

More precisely, assume that a packet sequence $(A,L,F)$ is fed into a FIFO system (\fref{fig-f1}). 
The following theorem establishes that the minimal interleaved regulator does not add anything to the worst-case delay of the FIFO system.

\begin{figure}[h]
 \includegraphics[width=0.50\textwidth]{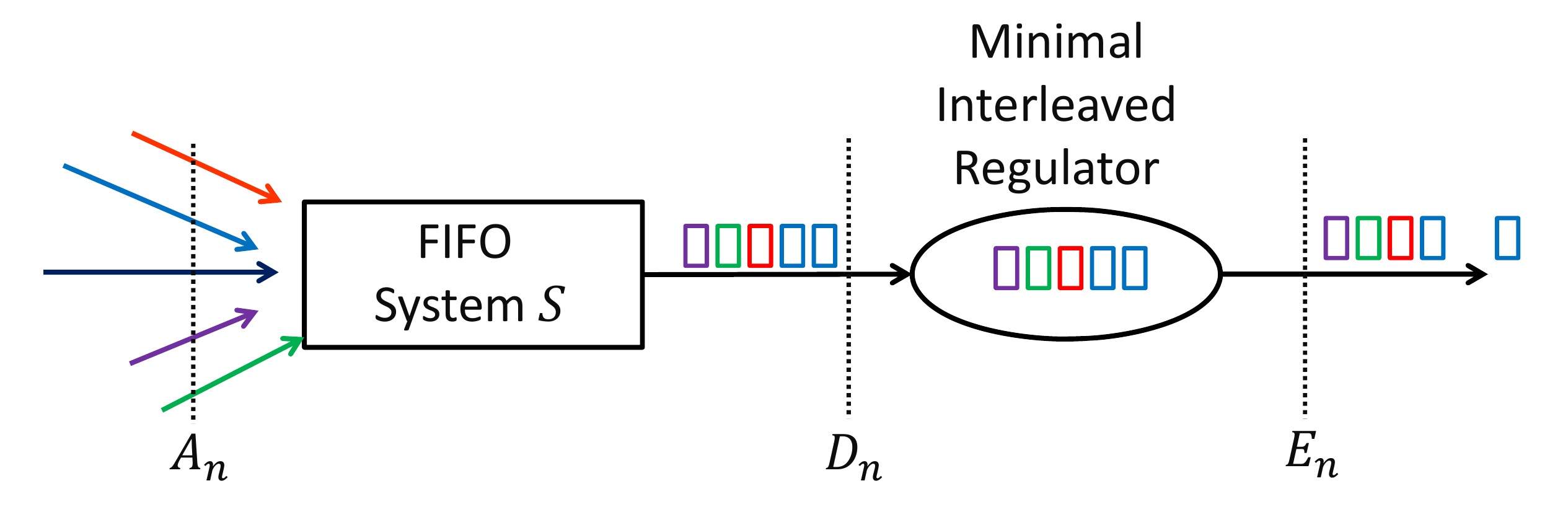}
\mycaption{Configuration for \thref{theo-ir}. }
\label{fig-f1}
\end{figure}

\begin{theorem}
\label{theo-ir}
  Assume that a packet sequence $(A,L,F)$ is fed into a FIFO system $S$. Assume that every input flow $f$ is $\Pi^f$ regular and that the operators $\Pi^f$ satisfy conditions C1-C4. The output packet sequence $(D,L,F)$ is fed into a minimal interleaved regulator with operator $\Pi^f$ for flow $f$; its output is the packet sequence $(E,L,F)$. The worst-case delay of the combination is the same as the worst-case delay of the FIFO system $S$ alone. In other words:
\be
\sup_{n\in\Nats^+}\lp D_n-A_n\rp = \sup_{n\in\Nats^+}\lp E_n-A_n\rp
\label{eq-sff}
\ee
\end{theorem}

\ntxt{\paragraph*{Proof}The proof mimicks the proof of \thref{theo-pfr}, and consists in comparing the minimal interleaved regulator with a damper with parameter equal to the worst-case delay of system $S$ over all flows.\cqfd}
\ntxt{
\subsection{Comparison With Per-Flow Minimal Regulation}
Concerning the effect on worst-case delay, we can compare the minimal interleaved regulator  with a bank of per-flow minimal regulators, as in \fref{fig-rpr}. 

\begin{figure}[h]
 \includegraphics[width=0.5\textwidth]{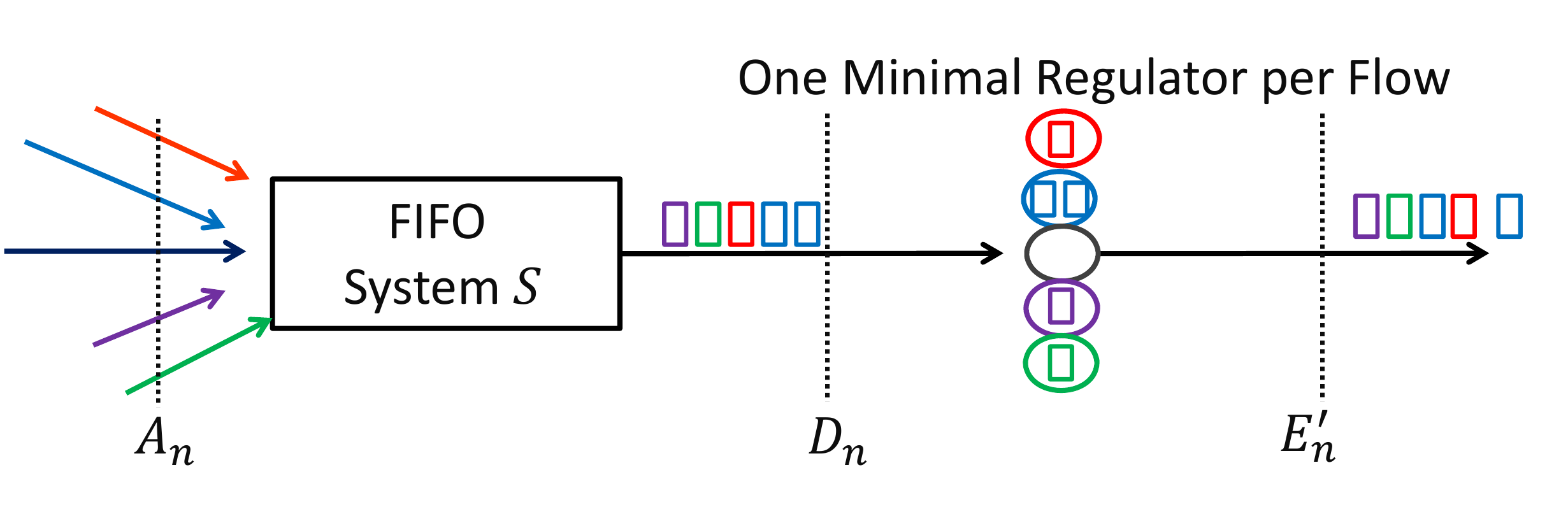}
\mycaption{Configuration for comparing interleaved regulator with a bank of per-flow regulators. }
\label{fig-rpr}
\end{figure}

Assume the same setting as in \thref{theo-ir}, with the difference that every flow is submitted to its own minimal regulator, with one regulator queue per flow. Let $E'$ be the sequence of output times for the bank of per-flow regulators.  The bank of per-flow regulators is not globally FIFO, since the head-of-the-line blocking of the minimal interleaved regulator does not happen here, i.e. the sequence $E'$ is not necessarily wide-sense increasing. For every flow $f$, let $d^f$ be its worst-case delay at system $S$. We can apply \thref{theo-pfr} to every flow $f$ in this configuration and obtain that
 \be
 \sup_{n\in\Nats^+}\left(E'_n-A_n\right)\ind{F_n=f}=d^f, \;\;\;\forall f
 \label{eq-pfdb}
 \ee
Compare with \thref{theo-ir}, with which there is a subtle difference. Indeed, \thref{theo-ir} states that the worst-case delay $d=\max_f d^f$ across all flows at system $S$ is not increased by the downstream minimal interleaved regulator. In some cases, the worst-case delay $d^f$ at system $S$ for some flow $f$ may be less than for other flows flows, and thus $d^f<d$ for some flows $f$. A typical case is when different flows have different maximum packet lengths and when the delay at $S$ includes a transmission delay proportional to packet length. Another case is when different flows require different processing delays. In such cases, the delay bound of \thref{theo-ir} gives
\be
 \sup_{n\in\Nats^+}\left(E_n-A_n\right)\ind{F_n=f}\leq d, \;\;\;\forall f
 \label{eq-pfdb2}
 \ee
which is weaker than the delay bound in \eref{eq-pfdb} obtained with per-flow regulators for all flows $f$ such that $d^f<d$. The example in Appendix~\ref{app-ex} shows that it is possible to have equality in \eref{eq-pfdb2} for a flow with $d^f<d$. In other words, interleaved regulation ``comes for free" only for the overall worst-case delay across all flows; for the flows that have a worst-case delay less than the overall worst-case, there might be an increase, up to the overall worst-case delay.


We can also easily establish a stronger, per-packet inequality.
Consider one flow of interest, say $f$, and observe only the packets of this flow as they go through the configuration in \fref{fig-f1}. The output sequence $(E^f, L^f)$ is $\Pi^f$-regular and the minimal interleaved regulator is FIFO for this flow (since it is globally FIFO). Therefore, the minimal interleaved regulator is also a per-flow regulator for flow $f$, though probably not minimal. It follows that $E_n\geq E'_n$ whenever packet $n$ belongs to flow $f$. Since this holds for every flow $f$, it follows that
 \be
  E_n\geq E'_n,\;\;\;\forall n
 \ee
In other words, the minimal interleaved regulator cannot beat the per-flow regulator. The example in Appendix~\ref{app-ex} shows that, in general, the above inequality is strict for a non empty set of packets.
}

\ntxt{
\subsection{Application to FIFO Per-Class Networks}
}

This result is quite important for networks that do per-class scheduling as it shows that it is possible to avoid the burstiness cascade while keeping only FIFO queues. Following \cite{specht2016urgency}, assume that we place one minimal interleaved regulator per switch input port and per traffic class, before the input to a queuing point, as illustrated in \fref{fig-tsn}.
The above theorem can then be applied, where the FIFO system $S$ is the upstream node that feeds the interleaved regulator. Note that in the usptream node, there are other flows that are not fed to the interleaved regulator. Thus, more precisely, the FIFO system $S$ can be defined as the system that transforms the multi-flow packet sequence \ntxt{consisting in} all packets of all flows that come from the upstream node.

\begin{figure}[h!]
\centering
 \includegraphics[width=0.50\textwidth]{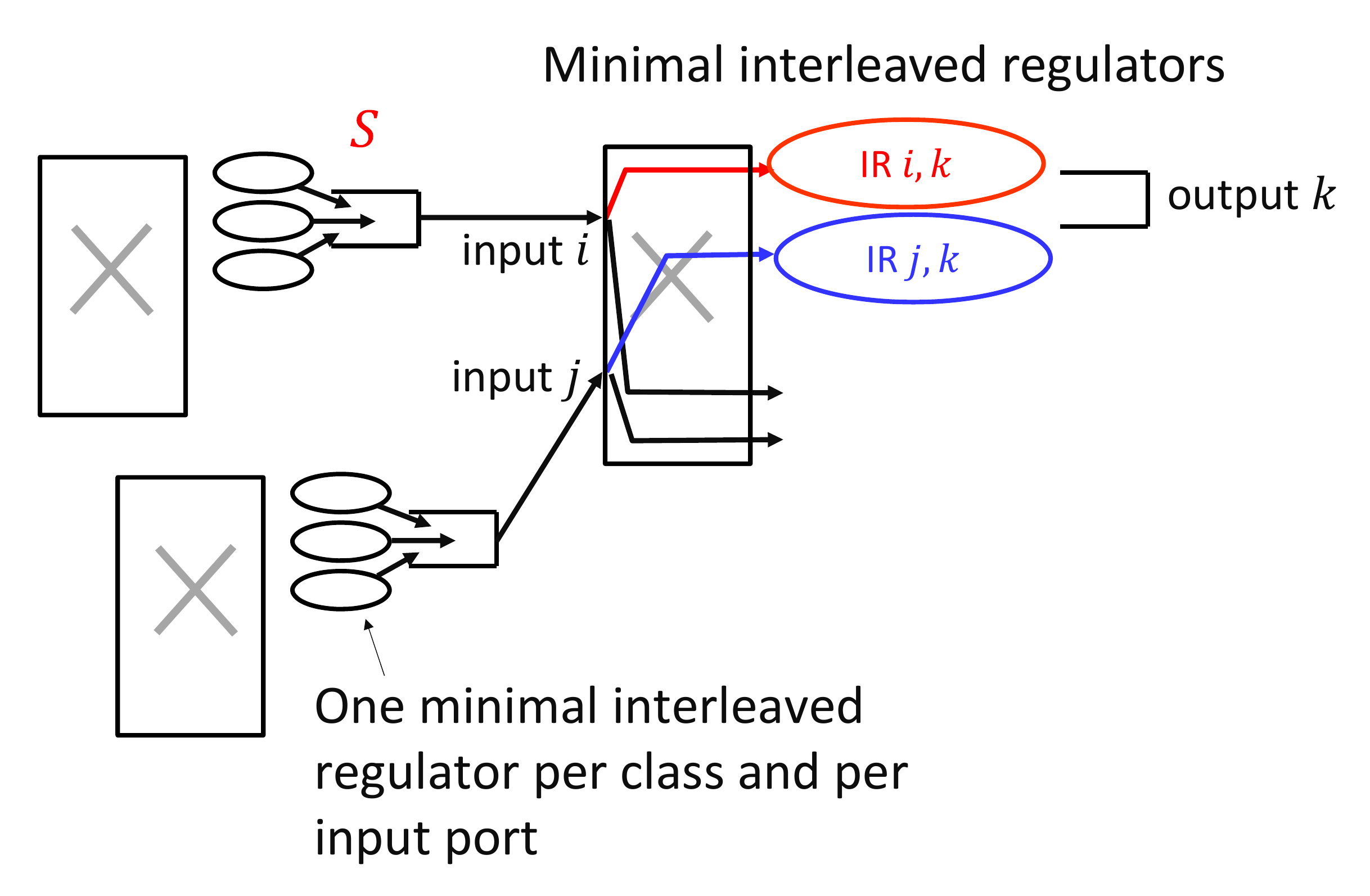}
\mycaption{Use of Interleaved Regulators as proposed by \cite{specht2016urgency}. \ntxt{Minimal interleaved regulators are inserted before the output port schedulers. The figure assumes that there is only one traffic class; at output port $k$ there is one minimal interleaved regulator (such as IR$i,k$ and IT$j,k$) per input port.}}
\label{fig-tsn}
\end{figure}

When computing end-to-end delay bounds, minimal interleaved regulators can then be ignored, since, by the above theorem, their delay can be absorbed into the previous node delay. \ntxt{This holds for \emph{any} valid delay bounds, because \thref{theo-ir} is about the worst-case delay.} Furthermore, since the input flows to any queuing point is the output of an interleaved regulator and thus is regulated, delay and backlog bounds can be computed \ntxt{using network calculus computations; see \cite{mohammadpour2018end2} for a detailed example.}

TSN assumes that every source satisfies what we call here the ``TSN packet rate regulation" rule in \eref{eq-prsc}. Therefore, in a TSN network\ntxt{,} it would \ntxt{be} consistent to use \ntxt{a minimal} interleaved \ntxt{regulator} where the regulation rule for every flow is ``TSN packet rate regulation" (rather than LRQ or leaky bucket, as is currently proposed by UBS). \ntxt{Observe however that the ``TSN packet rate regulation" is complex to implement in a regulator, as it requires remembering the dates at which the most recent packets were sent. A simpler alternative is to replace it with the ``packet burstiness" constraint (\sref{sec-lpr}), which requires storing only one counter per flow.}

%% file: conc.tex
\section{Conclusion}
\label{sec-conc}
Motivated by the Urgency Based Scheduler of \cite{specht2016urgency}, we have introduced a new theory of \ntxt{traffic regulators that is able to explain the ``reshaping-for free" property of minimal regulators, and can be extended to interleaved regulators, which handle multiple flows in a single FIFO queue}. This theory extends the existing, non compatible theories of g-regulators and arrival curve constraints, and also sheds some light on their relationship. It also gives a practical means to avoid burstiness cascades in per-class FIFO networks.

\ntxt{In future research, it might be interesting to explore the service guarantees offered by minimal regulators. Indeed the packetized greedy shaper, which is the minimal regulator for an arrival curve constraints, offers a service guarantee in the form of a service curve (a min-plus concept). Similarly, Chang's minimal g-regulator is a g-server (the max-plus counterpart of a service curve element). Since Pi-regulation subsumes both arrival curve constraints and g-regulation, it will be interesting to find ways of expressing the service guarantees offered by the minimal Pi-regulator and the minimal interleaved regulator.}


%% file: app.tex

\subsection{Proof of \lref{lem-leminv2}}

%
%

$\bullet$ $\Leftarrow$  is \eref{eq-invineq0}.

$\bullet$ $\Rightarrow$:
We have $t\geq f^{\downarrow}(x)$. If $t>f^{\downarrow}(x)$, the conclusion follows from \eref{eq-invineq}. Else we have $t=f^{\downarrow}(x)$. For any $s>t$, we have $f(s)\geq x$ by \eref{eq-invineq}.  Therefore $\limit{s}{t, s>t} f(s)\geq x$. Since $f()$ is right-continuous, $\limit{s}{t, s>t} f(s)=f(t)$.

\subsection{Proof of \lref{lem-inv}}
We do the proof for the statement $f^{\downarrow}=\lp f^+ \rp^{\downarrow}$. The proof for $f^{\uparrow}=\lp f^- \rp^{\uparrow}$ is exactly similar and is left to the reader.

Note that $f(t)\leq f^+(t)$ for all $t$ therefore
$\lc t: f(t)\geq x \rc \subseteq \lc t: f^+(t)\geq x \rc$, thus $f^{\downarrow}(x)\geq (f^+)^{\downarrow}(x)$.

Assume now, by contradiction, that $f^{\downarrow}(x)> (f^+)^{\downarrow}(x)$. For any $t\mst (f^+)^{\downarrow}(x)<t< f^{\downarrow}(x)$, we have:
 \begin{enumerate}
   \item $t< f^{\downarrow}(x)$ and since $f^{\downarrow}(x)$ is the infimum of the set $\lc t: f(t)\geq x\rc $, it follows that $t$ is not in this set, i.e. $f(t)<x$
   \item $t>(f^+)^{\downarrow}(x)$ and by \eref{eq-invineq} it follows that $f^+(t)\geq x$
 \end{enumerate}
 Thus, for any $t\in \lp (f^+)^{\downarrow}(x),\; f^{\downarrow}(x)\rp $ we have $f^+(t)>f(t)$ i.e. $t$ is a point of discontinuity of $f$. But this is impossible because this interval is not a countable set. Thus it is not possible that $f^{\downarrow}(x)> (f^+)^{\downarrow}(x)$, which proves that $f^{\downarrow}(x)= (f^+)^{\downarrow}(x)$.

\subsection{Proof of \lref{lem-inv-n}}
We do the proof for the statement ${(f^{\uparrow})}^-=f^{\downarrow}$. The proof for ${(f^{\downarrow})}^+=f^{\uparrow}$ is exactly similar and is left to the reader.

We can rewrite the definitions in \eref{eq-inv-l2} and \eref{eq-inv-u1} as
\bear f^{\downarrow}(y)&=&
\sup_{s\in\Reals^+}\lp s \ind{f(s)> y}\rp\label{eq-inv-l2b}\\
f^{\uparrow}(y)&=&
\sup_{s\in\Reals^+}\lp s \ind{f(s)\leq y}\rp
\eear
Therefore, using associativity of $\sup$:
\bear
(f^{\uparrow})^-(x)&\eqdef&\sup_{0\leq y<x}f^{\uparrow}(y)\\
&=&\sup_{0\leq y<x}\lp \sup_{s\in\Reals^+}\lp s \ind{f(s)\leq y}\rp\rp\\
&=&\sup_{s\in\Reals^+}\lp \sup_{0\leq y<x}\lp s \ind{f(s)\leq y}\rp\rp\\
&=&\sup_{s\in\Reals^+}\lp s\; \varphi(s,x)\rp
\eear with $\varphi(s,x)\eqdef\sup_{0\leq y<x}\lp \ind{f(s)\leq y}\rp$.
Now if $x>f(s)$ then $\varphi(s,x)=1$ and if $x\leq f(s)$ then $\varphi(s,x)=0$. Therefore
$\varphi(s,x)=\ind{x> f(s)}$. Thus

\be
(f^{\uparrow})^-(x) = \sup_{s\in\Reals^+}\lp s\;\ind{x> f(s)}  \rp=f^{\downarrow}(x)
\ee where the last equality is by \eref{eq-inv-l2b}.

\subsection{Proof of \thref{theo-mpmpgs}}

$\bullet$ 1)$\Rightarrow$ 3):

Consider some packet numbers $1\leq m\leq n$. If $m=n$ then 3) trivially holds because $\sigma^{\downarrow}(0)=0$. Assume now that $m<n$. Let $T()$ be the arrival time function defined by \eref{eq-deft}. Take $y=L_1+...+L_n - \epsilon$ with $0<\epsilon\geq L_n$ and $x=L_1+...+L_m$. Because $L_n$ is integer, we have
$T(y)=A_n$ and $T(x)=A_m$.

By \cite{liebeherr2017duality}, the max-plus traffic envelope condition \eref{eq-mpte} also holds. Thus
\be
 A_n-A_m= T(y)-T(x)\geq \sigma^{\uparrow}(y-x)=\sigma^{\uparrow}(L_m+...+L_n-\epsilon)
\ee Take the limit of the above equation as $\epsilon\to 0$ and obtain
\be
A_n-A_m\geq (\sigma^{\uparrow})^-((L_m+...+L_n)
\ee

By \lref{lem-inv-n}, $(\sigma^{\uparrow})^-=\sigma^{\downarrow}$,  which concludes this part of the proof.

~\\

$\bullet$ 3)$\Rightarrow$ 2):

Consider some packet numbers $1\leq m\leq n$. If $m=n$ then 2) trivially holds because $\sigma(0)\geq 0$ and $\sigma^+(0)\geq 0$. Assume now that $m<n$. \eref{eq-mpmpg2} follows from \lref{lem-leminv2} applied to $f()=\sigma^+()$.

~\\

$\bullet$ 2)$\Rightarrow$ 1):

\emph{Part 1: no simultaneous arrivals.}

We first prove this case assuming that there cannot be simultaneous arrivals, namely we assume $A_n<A_{n+1}$ for all $n\in \Nats^+$.

Consider $s, t\in \Reals^+$ with $0\leq s\leq t$. If $s=t$ then \eref{eq-ac} trivially holds. Assume therefore that $0\leq s<t$. Let $\calA=\lc A_1, A_2, ...\rc$. We consider several cases:

\par\noindent Case 1:  $\calA \cap [s, t]$ is empty. In this case. $R(t)-R(s)=0$ and \eref{eq-ac} is trivially satisfied.
~\\

\par\noindent Case 2: $\calA \cap [s, t]$ is nonempty, $s\nin \calA$  and $t\nin \calA$. Let $m$ be the smallest packet number such that $s< A_m$ and let $n$ be the largest packet number such that $A_n<t$, so that $\calA \cap [s, A_m)$ and $\calA \cap (A_n, t]$ are empty. Therefore
 \be
 R(t)-R(s)=\sum_{j=m}^{n}L_j
 \ee
and by \eref{eq-mpmpg2}
 \be
 R(t)-R(s)\leq \sigma^+(A_n-A_m)
 \ee

 We must also have
\be
s<A_m \leq A_n<t
\ee and thus $A_n-A_m< t-s$; by \eref{eq-fplus2}
\be
\sigma^+(A_n-A_m)\leq \sigma(t-s)
\ee
This concludes the proof in this case.
~\\

\par\noindent Case 3:
$s\in\calA$ and $t \nin\calA$. Thus $s=A_m$ for some $m$. Let $n$ be the largest packet number such that $A_n<t$. We have therefore
\be
R(t)-R(s)=\sum_{j=m}^{n}L_j
\ee
and
\be s=A_m\leq A_n<t\ee
The rest of the proof in this case is as in Case 2.
~\\

\par\noindent Case 4:
$s\nin\calA$ and $t \in\calA$. Thus $t=A_n$ for some $n$. Let $m$ be the smallest packet number such that $s<A_m$. We have therefore
\be
R(t)-R(s)=\sum_{j=m}^{n-1}L_j
\ee
and
\be s<A_m\leq A_n=t\ee
Thus
 \be
 R(t)-R(s)\leq \sum_{j=m}^{n}L_j\leq \sigma^+(A_n-A_m)
 \ee
where the last inequality is by \eref{eq-mpmpg2}. Now
\be A_n-A_m < t-s
\ee
 \be
 R(t)-R(s)\leq \sigma(A_n-A_m)
 \ee
Now $A_n-A_m\leq t-s$ and $\sigma()$ is wide-sense increasing, thus $\sigma(A_n-A_m)\leq \sigma(t-s)$, which concludes the proof in this case.
~\\

\par\noindent Case 5:
$s\in\calA$ and $t \in\calA$. Thus $s=A_m$ and $t=A_n$ for some $m\leq n$. We have therefore
\be
R(t)-R(s)=\sum_{j=m}^{n-1}L_j
\ee
If $m=n$ then $R(t)-R(s)=0$ and \eref{eq-ac} is trivially verified. We can therefore assume $m<n$. It follows that $0 \leq A_{n-1}-A_m<A_n-A_m$ therefore
\be\sigma^+(A_{n-1}-A_m)\leq \sigma(A_n-A_m)\ee
By \eref{eq-mpmpg2}
 \be
 R(t)-R(s)\leq \sigma^+(A_{n-1}-A_m)\leq \sigma(A_n-A_m)
 \ee

\emph{Part 2: with simultaneous arrivals.}

We now allow simultaneous arrivals in the flow $(A,L)$. We assume that there is a finite number of packet arrivals in every bounded interval. Indeed, if this does not hold, the conditions in the theorem are false and the equivalence holds.

We derive from $(A, L)$ another packet sequence, $(A',L')$ obtained by aggregating all packets that arrive at the same time under $(A,L)$. Formally, $(A',L')$ is defined by:
\beln
A'_1 =\min \lc A_m, m\in \Nats^+ \rc\\
A'_n = \min \lc A_m, m\in \Nats^+, A_m>A'_{n-1}\rc\\
L'_n=\sum_{j\in \Nats^+} L_j \ind{A_j=A'_n}
\eeln
Note that $L'_n$ is finite for every $n$ by our assumption and $L'_n$ is the sum of all packet sizes of all packets that arrive at the same instant. Note that $A'\in \calF_{inc}$ and there are no simultaneous arrivals in the flow $(A',L')$.

We next show that, for $i=1,2$, condition $i$ of the theorem holds for $(A,L)$ if and only if it holds for $(A',L')$, which will conclude the proof.

\par Condition 1): The cumulative arrival function $R()$ is the same for $(A,L)$ and $(A',L')$ so Condition $1$ of the theorem holds for $(A,L)$ if and only if it holds for $(A',L')$.

\par Condition 2):
Assume first that Condition 2) holds for $(A, L)$. Consider some fixed $m',n'\in\Nats^+$ and let $m$ be the index of the first packet such that $A_m=A'_{m'}$ and $n$ the index of the last packet such that $A_n=A'_{n'}$. We have
\be \sum_{j=m'}^{n'} L'_j = \sum_{i=m}^{n} L_i
\ee
and $A_n-A_m=A'_{n'}-A'_{m'}$. Since 
\eref{eq-mpmpg2} holds for $(A, L)$, it follows that it also holds for $(A',L')$.

Conversely, assume that Condition 2) holds for $(A', L')$ and consider some fixed $m,n\in\Nats^+$. Let Define and $m',n'$ by $A'_{m'}=A_m$ and $A'_{n'}=A_n$. We have
\be \sum_{j=m}^n L_j\leq \sum_{i=m'}^{n'} L'_i
\ee
and $A_n-A_m=A'_{n'}-A'_{m'}$. Since 
\eref{eq-mpmpg2} holds for $(A', L')$, it follows that it also holds for $(A,L)$.

This concludes the proof in this case.
~\\


\subsection{Proof of \thref{theo-minreg}}

1) We first prove that the system defined by $D_1=A_1$ and \eref{eq-defminreg} is a $\Pi$-regulator. We obviously have $D_n\geq D_{n-1}$ i.e. $D\in \calF_{inc}$ and $D_n\geq A_n$ for all $n \in  \Nats\ntxt{^+} $ thus this is a FIFO system.  Also $D_n\geq \Pi(D, L)_n$ by construction.

2) Next, we show by induction that $D'_n \geq D_n$.

Base Step: We have $D_1=A_1\leq D'_1$ because the $\Pi$-regulator is a FIFO system.

Induction Step: Assume $D'_m\geq D_m$ for $1\leq m\leq n-1$. Let $\bar{D}$ be the sequence defined by $\bar{D}_m=D'_m$ for $1\leq m\leq n-1$ and $\bar{D}_m=-\infty$ for $m\geq n$. We have
$D'\geq \bar{D}$ by induction hypothesis and by Condition C4, $\Pi(D',L)\geq \Pi(\bar{D},L)$. By Condition C2, $\Pi(\bar{D},L)_n=\Pi(\ntxt{D'},L)_n$. Therefore $\Pi(D',L)_n\geq \Pi(D,L)_n$. But since $D'$ is $\Pi$-regular, we also have $D'_n\geq \Pi(D',L)_n$. Thus
\be D'_n\geq \Pi(D,L)_n \ee
Now \be D'_n\geq D'_{n-1}\geq D_{n-1} \mand D'_n\geq A_n\ee because the $\Pi$-regulator is a FIFO system. Combining the last two inequalities gives
\be
D'_n\geq \max\lc A_n, D_{n-1},\Pi(D,L)_{n}\rc=D_n
\ee

3) If $D=A$ then since $D$ is $\Pi$-regular by 1) obviously $A$ is $\Pi$-regular.

Conversely, if $(A,L)$ is $\Pi$-regular then the identity system, which maps $(A,L)$ into itself, is a $\Pi$-regulator for this flow. By item 2), we have $D\leq A$. But since $D\geq A$ by construction it follows that $D=A$.

\subsection{Proof of \thref{theo-ir-min}}
1) We first prove that the system defined by $D_1=A_1$ and \eref{eq-defminir} is an interleaved regulator. We obviously have $D_n\geq D_{n-1}$ i.e. $D\in \calF_{inc}$ and $D_n\geq A_n$ for all $n \in  \Nats\ntxt{^+} $ thus this is a FIFO system.  Also by construction
\be D_n \geq \Pi^{F_n} \lp D^{F_n}, L^{F_n}\rp_{I(n)}\ee which is the same as
\be D^{F_n}_{I(n)} \geq \Pi^{F_n} \lp D^{F_n}, L^{F_n}\rp_{I(n)}\ee
which shows that $D^f\geq\Pi^f \lp D^{f}, L^f\rp$ for every flow $f$, i.e. every flow at the output is $\Pi^f$-regular.

2) Next, we show by induction that $D'_n \geq D_n$.

Base Step: We have $D_1=A_1\leq D'_1$ because the interleaved regulator is a FIFO system.

Induction Step: Assume $D'_m\geq D_m$ for $1\leq m\leq n-1$. Let $\bar{D}$ be the sequence defined by $\bar{D}_m=D'_m$ for $1\leq m\leq n-1$ and $\bar{D}_m=-\infty$ for $m\geq n$. We have
$D'\geq \bar{D}$ by induction hypothesis and thus $D'^{F_n}\geq\bar{D}^{F_n}$. By Condition C4, $\Pi^{F_n}(D'^{F_n},L^{F_n})\geq \Pi^{F_n}(\bar{D}^{F_n},L^{F_n})$. By Condition C2, $\Pi^{F_n}(\bar{D}^{F_n},L^{F_n})_{I(n)}=\Pi^{F_n}(D^{F_n},L^{F_n})_{I(n)}$. Therefore $\Pi^{F_n}(D'^{F_n},L^{F_n})_{I(n)}\geq \Pi^{F_n}(D^{F_n},L^{F_n})_{I(n)}$. But since $D'$ is $\Pi^{F_n}$-regular, we also have $D'_n=(D'^{F_n})_{I(n)}\geq \Pi^{F_n}(D'^{F_n},L^{F_n})_{I(n)}$.
Thus
\be D'_n\geq \Pi^{F_n}(D^{F_n},L^{F_n})_{I(n)} \ee
Now \be D'_n\geq D'_{n-1}\geq D_{n-1} \mand D'_n\geq A_n\ee because the interleaved regulator is a FIFO system. Combining the last two inequalities gives
\be
D'_n\geq \max\lc A_n, D_{n-1},\Pi^{F_n}(D^{F_n},L^{F_n})_{I(n)}\rc=D_n
\ee

3) Since the system is an interleaved regulator by item 1, every flow $f$ in the output sequence is $\Pi^f$ regular. Thus, if $D=A$, the input is equal to the output and every flow $f$ in the input sequence is also $\Pi^f$ regular.

Conversely, if every flow in $(A,L, F)$ is $\Pi^f$-regular then the identity system, which maps $(A,L, F)$ into itself, is an interleaved regulator for this packet sequence. By item 2), we have $D\leq A$. But since $D\geq A$ by construction it follows that $D=A$.

\ntxt{
\subsection{A Numerical Example}
\label{app-ex}
We choose per unit values where 1 data unit $=1200$~bytes and 1 time unit $=12\mu$sec.  Consider a scenario with 2 flows. For flow 1, all packets have length equal to 2 data units, and for flow 2 it is 1 data unit. Flow~1 is subject to a packet spacing regulation with $\tau_1=5$~time units and flow~2 to a packet spacing regulation with $\tau_2=10$~time units.

We use the notation of Figures~\ref{fig-f1} and \ref{fig-rpr}. The input packet sequence to the FIFO systems $S$ is $(A,L,F)$ with
\beln
 A=(0,5,5,10,15,15,10,25,25,...)\\
 L=(2,2,1,2,2,1,2,2,1,...)\\
 F=(1,1,2,1,1,2,1,1,2,...)
 \eeln
 In other words, flow 1 sends one packet every 5 time units and flow 2 sends one packet every 10 time units, which arrives immediately after an even-numbered packets of flow 1. At the input, both flows are conforming to their regulation constraints.

The output of the FIFO system $S$ is $(D,L,F)$ with
\beln
 D=(5,7,8,15, 17, 18, 25, 27, 28, ...)
 \eeln
 i.e., the odd packets of flow 1 have a response time of $5$ time units, the even packets of 2 time units, and the packets of flow 2 have a response time of 3 time units. This corresponds to the case where $S$ is a simple priority queue, where flows 1 and 2 are served with low priority at a rate equal to 1~p.u. (i.e. 800~Mb/s) and where a high priority packet preempts the server during time intervals $[0;3], [10;13], ...$. The worst-case delay at $S$ for flow 1 is $d^1=5$~time units and for flow 2 it is $d^2=3$~time units. The overall worst-case delay at $S$ is $d=5$~time units.

 The output of the minimal interleaved regulator is  $(E,L,F)$ with
\beln
 E=(5,10,10,15,20, 20, 25, 30, 30, ...)
 \eeln
 Indeed the odd packets of flow~1 are not delayed by the minimal interleaved regulator, but the even packets are delayed because they arrive too early with respect to a spacing constraint of 5~time units. Packets of flow~2 are delayed because they stand behind the even packets of flow~1. For flow~1, the worst-case delay at the combination of $S$ and the interleaved regulator is $d^1_{tot}=5$~time units and for flow 2 it is $d^2_{tot}=5$~time units as well. The overall worst-case delay is $d_{tot}=d=5$~time units. The overall worst-case delay is not increased by the minimal interleaved regulator, but the worst-case delay of flow~2 \emph{is} increased.

 The output of the bank of per-flow regulators is  $(E',L,F)$ with
\beln
 E'=(5,10,8,15,20, 18, 25, 30, 28, ...)
 \eeln
 The difference with the interleaved regulator is that packets of flow~2 are not delayed (as a result, the bank of per-flow regulators is not globally FIFO). For packets of flow~2 we have $E'_n<E_n$.
} 